\documentclass[runningheads]{llncs}
\usepackage{graphicx}
\usepackage[nolist]{acronym}
\usepackage{dirtytalk}
\usepackage{color}
\usepackage{listings,chngcntr}
\usepackage{listings-rust}
\usepackage[hyphens]{url}
\usepackage{hyperref}
\usepackage[hyphenbreaks]{breakurl}
\usepackage{orcidlink}

\usepackage{tikz}
\usetikzlibrary{arrows,arrows.meta,calc,decorations,decorations.pathmorphing,decorations.pathreplacing,fit,positioning,shadows,shapes,shapes.multipart}

\tikzset{
	token/.style={text height=3mm},
	node distance = 1em
}

\makeatletter
\newcommand{\printfnsymbol}[1]{%
  \textsuperscript{\@fnsymbol{#1}}%
}
\makeatother

\newcommand{\citerustref}[1]{\footnote{\url{https://doc.rust-lang.org/1.79.0/reference/ #1}}~\cite{rust_lang_ref}}

\begin{document}

\counterwithin{lstlisting}{section} 

\title{Towards Modified Condition/Decision Coverage of Rust}
%
%
\author{Wanja Zaeske\inst{1}\thanks{equal contribution}\orcidlink{0000-0002-1427-2627} \and
Pietro Albini\inst{2}\printfnsymbol{1}\orcidlink{0009-0003-3863-1959} \and
Florian Gilcher\inst{2}\orcidlink{0009-0002-7870-7359} \and
Umut Durak\inst{1}\orcidlink{0000-0002-2928-1710}}
\authorrunning{W. Zaeske et al.}
%
\institute{German Aerospace Center, Institute of Flight Systems, Lilienthalpl. 7, 38108 Braunschweig, Germany\email{\{wanja.zaeske,umut.durak\}@dlr.de}\\
\url{http://dlr.de/ft/en/ssy} \and
Ferrous Systems GmbH, Boxhagener Straße 79, 10245 Berlin, Germany\\
\email{\{pietro.albini,florian.gilcher\}@ferrous-systems.com}\\
\url{https://ferrous-systems.com/}}
\maketitle              
\begin{abstract}
Testing is an essential tool to assure software, especially so in safety-critical applications.
To quantify how thoroughly a software item has been tested, a test coverage metric is required.
Maybe the strictest such metric known in the safety critical systems is \ac{MCDC}, which DO-178C prescribes for the highest software assurance level in aviation.
In the past, ambiguities in the interpretation of \ac{MCDC} have been resolved already, i. e. in CAST-10.
However, some central features of the Rust programming language necessitate further clarification.
This work investigates aforementioned features, in particular \textit{pattern matching}, providing a consistent view on how to apply \ac{MCDC} to Rust.
Hence, this paper informs the implementation of Rust \ac{MCDC} tools, paving the road towards Rust in high-assurance applications.
\keywords{\ac{MCDC} \and Rust \and DO-178C}
\end{abstract}
%
%
%
%
\acresetall
\begin{acronym}[TDMA]
    \acro{API}{Application Programming Interface}
    \acro{AST}{Abstract Syntax Tree}
    \acro{DC}{Decision Coverage}
    \acro{MCDC}[MC/DC]{Modified Condition/Decision Coverage}
    \acro{CAST}{Certification Authorities Software Team}
    \acro{TQL}{Tool Qualification Level}
\end{acronym}
\section{Introduction}

\begingroup
\let\thefootnote\relax
\footnotetext[0]{This paper refers to Rust 1.79, later versions are expected to progress past the described status.}
\endgroup

Over the past decade, Rust developed into a prominent systems programming language.
Together with its user-base grew the desire to apply Rust to high assurance, safety-critical systems.
However, the Rust ecosystem is not yet ready to serve these domains fully.
The lack of a \ac{MCDC} tool for Rust is one of the reasons precluding Rust from some use-cases~\cite{rust_code_analysis_bachelor_thesis}.
For example, DO-178C, the standard on software in avionics, mandates \ac{MCDC} in Level A systems (the highest criticality level known to the aviation sector)~\cite{do178c}.
Level B systems still require \ac{DC}, for which there is no Rust compatible tool out there yet, either.

One of the reasons for the current lack of tooling is the significant complexity introduced by Rust's expressive syntax.
In particular, Rust's pattern matching comprises hidden conditions, which are not covered by the specific wording on \ac{DC} and \ac{MCDC} found in DO-178C, yet most definitely should be considered as decisions.
Further on, Rust's const by default introduces two kinds of constants: \lstinline|let| bound non-mutable variables and actual \lstinline|const| declared constants.
As the clarification about \ac{MCDC} in the CAST-10 position paper introduces a more lax treatment of \ac{MCDC} around constants, these two kinds require further elaboration as well~\cite{cast10}.
Other than that, the implementation of macros in Rust hinders a common way of measuring \ac{MCDC} by source code instrumentation, for example described in \cite{smother}.
Smaller issues, like the control flow diverting \lstinline|?| operator or the nesting of conditional statements (which itself may become expressions) might affect \ac{MCDC} as well, thus requiring brief discussion too.

These topics are structured as following in the paper:
\autoref{sec:background} first introduces the motivation for and definition of \ac{MCDC}.
This is followed by an explanation of Rust's pattern matching and its distinction from C/C++'s switch case statement.
The chapter closes with a contemplation of Rust's macros impeding source code instrumentation.
\autoref{sec:role_of_pattern_matching_in_rust} justifies the utilization of pattern-matching in safety-critical software.
Then \autoref{sec:mcdc-for-pattern-matching} thoroughly discusses the matter of pattern-matching for \ac{MCDC} while introducing terms to accurately describe patterns and their inherent decisions.
A brief discussion of the remaining issues is laid out in \autoref{sec:mcdc-further-considerations}.
While there is little literature on \ac{MCDC} for programming languages with pattern-matching, \autoref{sec:related_works}
summarizes previous works with similar goals.
Finally, \autoref{sec:certification} discusses matters of certification, focusing on the \ac{TQL} that a Rust \ac{MCDC} tool has to reach, the associated effort and consequently the feasibility of measuring \ac{MCDC} for software written in Rust.
Closing in, \autoref{sec:conclusion} concludes the aforementioned content while sketching the current path of \ac{MCDC} capability in Rust by the Rust Project.

\section{Background}
\label{sec:background}

Testing is an essential part of software assurance.
However, while testing can reveal errors, it can not prove the absence of errors.
But, the degree to which a test exercises the software can be measured and analyzed.
This is described as test coverage, and various test coverage metrics have been established.
While tests serve to verify software, test coverage metrics are used to verify the tests themselves.
A simple example is statement coverage: \say{Every statement in the program has been invoked at least 
once}~\cite{do178c}.
By requiring the above, some confidence can be derived from a set of tests; now it is known that the tests at least cause every statement in the code to be executed.
Major error classes are guaranteed to be found by this, e.g. code, which unconditionally crashes the program, will be uncovered if statement coverage is achieved.

However, many errors are conditional, and not all errors manifest in program crashes.
Code that expresses logic in form of boolean algebra might not cause a program crash, but if erroneous, can still cause system behavior that results in an aircraft crash.
Statement coverage guarantees that all statements were executed with \textit{some} values, but they are not guaranteed to identify problems with \textit{all} values that the statements operate on.

To gain confidence that the software does not exert unintended behavior, a stricter test coverage metric is required.
An example for that would be \ac{DC}: \say{Every point of entry and exit in the program has been invoked at least once and every decision in the program has taken all possible outcomes at least once}~\cite{do178c}.
By requiring \ac{DC}, more error-paths can be discovered: conditional faults depending on the outcome of boolean algebra are guaranteed to occur during testing, and hence have a good chance of being found as well.

However, \ac{DC} still does not exercise the actual logic of decisions, once both true and false results from a decision were observed \ac{DC} is satisfied.
Here \ac{MCDC} comes into play: on top of \ac{DC}, it requires each condition in a decision to have an effect on the decision's outcome.
That is, there is a fixed value assignment to all other conditions, so that toggling the one remaining condition also toggles the decision's outcome~\footnote{Minor variations of this exist, DOT/FAA/AR-01/18 discusses these in detail~\cite{ar0118}}.
Of the aforementioned test coverage metrics, \ac{MCDC} requires the most exhaustive set of tests for decision logic.

\subsection{Modified Condition/Decision Coverage}

In the previous paragraphs the terms \textit{condition} and \textit{decision} have been used without a proper explanation.
As the demand for \ac{MCDC} in avionic software arises from DO-178C, it makes sense to review its definition of the terms.
The atomic unit for \ac{MCDC} is the condition, an expression that evaluates to a boolean without itself containing boolean operators (except for the unary \texttt{not} operator, which is allowed in conditions)~\cite{do178c}.
From there, a new definition is derived: the decision.
A decision itself is an expression that evaluates to a boolean, that however may include zero or more boolean operators.
It follows that each condition as well is a decision, and that multiple conditions can be combined into a decision using boolean operators.
Now, \ac{MCDC} as per DO-178C is achieved, if:

\begin{itemize}
    \item All entry- and exit-points have been invoked at least once
    \item Every condition in each decision was both true and false at least once
    \item Each decision evaluated both to true and to false at least once
    \item Every condition in a given decision has been shown to independently affect the decisions outcome
\end{itemize}~\cite{do178c,cast10,nasa_mcdc_tut}

Satisfying the criteria mentioned above requires for a thorough analysis of the source code to find all conditions and decisions, and subsequently for an elaborate test suite.
During testing, instrumentation can be used to verify that all criteria of \ac{MCDC} are satisfied.
As all tests must be traceable to requirements~\cite{do178c}, \ac{MCDC} ensures that all possible implications of all boolean logic in the software are tested to comply with the requirements.

There has been some misunderstanding around the meaning of decision, where some claim that a decision may only be present at a branch point.
Now the \ac{CAST} was an international group of certification and regulatory authority representatives which authored recommendations in the form of position papers~\cite{faa_cast_website}.
And in their report CAST-10~\cite{cast10} they highlight, that this interpretation (branch points only) of decision alleviates the exhaustiveness of \ac{MCDC}.
Thus this interpretation is to be refused, and the literal, wider interpretation of decision as described in DO-178C is preferred instead.
Similar loopholes to evade the literal definition of condition, like expression of boolean logic through bit-wise logic and integer arithmetic, are as well discouraged and intolerable for any safety critical software.~\cite{cast10}

\subsection{Switch Case in C/C++}
\lstset{language=C}

The closest, yet far less complicated relative to pattern-matching in C or C++ are switch-case statements.
Hence, considerations specific to switch-case statements may guide the understanding of pattern-matching for \ac{MCDC}.
Unfortunately, the \ac{MCDC} literature is rather sparse with regard to switch-statements.

CAST-10 only refers to them while characterizing the more lax, eventually dismissed definition of decisions as \say{branch points}~\cite{do178c,cast10}.
The NASA tutorial on \ac{MCDC} briefly advises: \say{Note that a case statement may be handled similarly to the if-then-else statement}~\cite{nasa_mcdc_tut}.
Other document such as DO-178C or DOT/FAA/AR-01/18 do not elaborate at all on the handling of switch-statements~\cite{do178c,ar0118}.

Switch-statements are interesting, as they do not feature a directly visible decision (similar to pattern-matching).
Instead, the hidden decision's conditions are pulled apart: one expression after the \lstinline|switch| keyword, and one expression following each \lstinline|case| keyword.
The hidden decision then contains the test for equality of both expressions.
C and C++ both mandate the condition in a switch-statement and the expressions after \lstinline|case| to be of integer type (or in the case of C++ to be convertible to integer type)~\footnote{Enum types are allowed as well, however, they can be lowered to integer types}~\cite{ISO9899,ISO14882}.
Hence in these two languages, switch-statements resemble a single decision per case.
If control flow reaches a \lstinline|case| which is not terminated by a \lstinline|break| statement, then the control flow continues with the following case~\cite{ISO9899,ISO14882}.
Therefore, each case's decision evaluates to true even if only the previous case's evaluated to true, provided that the previous case was not terminated via a \lstinline|break|.

Equipped with this knowledge, switch-statements pose no ambiguity for the application of \ac{MCDC} in C or C++:
they can be converted to series of if/else statements, each of which containing one decision (except for the unconditional \lstinline|default| case).

\lstset{language=Rust}

\subsection{Enums in Rust}
\label{sec:enums_in_rust}

The Rust programming language features a powerful syntax, that significantly expands up on widely used systems programming languages like C.
One feature of particular interest for this paper are so called sum types, which are denoted using the \lstinline|enum| keyword.
They form a primary (but not the sole) reason for the existence of pattern-matching in Rust, the major concern of this paper.
This section hence introduces them, to motivate a typical use-case of Rust's pattern-matching.

Compared to enums in C, Rust also allows for the different variants of an enum to contain internal values of different types, like \lstinline|unions| in C.
Unlike C, Rust however guards the programmer against using the value in an enum variant as the wrong data type.
The programmer has to (pattern-) match a Rust enum instance against the specific variant containing a value, before being allowed to use that value (see \autoref{lst:enum_match} for an example).

\subsection{Pattern Matching in Rust}
\label{sec:pattern_matching_intro}

The polymorphic nature of enums in Rust demands for some elaborate syntax to maintain programming ergonomics.
After all, a Rust enum can contain any number of values, each of which may be of a different or the same type, and only at runtime may it be known which variant a variable holds.
The core syntax to achieve this was already demonstrated in \autoref{lst:enum_match}: \say{pattern matching}~\footnote{\url{https://doc.rust-lang.org/book/ch18-00-patterns.html}}\cite{rust_lang_ref}, similar to and inspired by OCaml's pattern matching~\footnote{\url{https://ocaml.org/docs/basic-data-types}}.

Pattern-matching enables two effects.
On one hand, pattern-matching allows value-driven branching.
For example, only if an enum value is of a certain variant (see \autoref{lst:enum_match}) or when a slice is of a given length (see \autoref{lst:complex_pattern}) will a certain code block be executed.
On the other hand, pattern matching also allows to bind expressions to values, possibly extracting inner fields of nested data types.
\autoref{lst:enum_match} is a concise example of this, the second and third \lstinline|match| arm bind the inner value of \lstinline|Passenger(u16)| to the variable \lstinline|n|.
Of the two effects, the former is particularly interesting to \ac{MCDC}: a series of hidden comparisons for equality decides how the program branches.
This is, in other words, a hidden decision comprised of multiple conditions.

From an economical standpoint, pattern-matching lets the programmer express the expected shape of data and inner fields they want to access using a syntax closely resembling the initialization of values, and it is up to the compiler to convert the pattern into the appropriate sequence of conditions and branches.

Pattern matching is used in many of Rust's syntactical constructs, like \lstinline|match|, \lstinline|if let|, \lstinline|let else|, \say{\lstinline|let| chains}, and the \lstinline|matches!| macro, and is pervasive across Rust programs due to its versatility.

\begin{lstlisting}[label=lst:enum_match,caption=Pattern matching of an enum]
enum Person { Crew, Passenger(u16) }
let person = Person::Passenger(3);
match person {
    Person::Crew => print!("crew member"),
    Person::Passenger(n @ ..=8) => print!("vip, seat {n}"),
    Person::Passenger(n) => print!("passenger, seat {n}")
}
\end{lstlisting}

As an example, \autoref{lst:enum_match} uses a \lstinline|match| expression to check the content of the \lstinline|person| variable.
If the enum variant is \lstinline|Person::Crew|, then the first match arm is executed.
Else, the enum variant must be \lstinline|Person::Passenger|.
Now a distinction is made: if the inner \lstinline|u16| in \lstinline|Person::Passenger| falls in the inclusive range literal $\left[ -\infty, 8 \right]$, then it is bound to the variable \lstinline|n| and the second match arm is executed.
Otherwise the inner value is bound to \lstinline|n| before executing the third match arm.
Patterns can also get more complex, checking complex data structures:

\begin{lstlisting}[label=lst:complex_pattern,caption={Matching of a complex pattern with \lstinline|if let|}]
let values: &[Option<i32>] = &[Some(1), None, Some(2)];
if let [Some(first), None, Some(1..)] = values {
    println!("First value is {first}");
}
\end{lstlisting}

In \autoref{lst:complex_pattern}, the single pattern in the \lstinline|if let| expression is matching that the slice has a length of 3, the first value is \lstinline|Option::Some|, the second value is \lstinline|Option::None|, the third value is \lstinline|Option::Some|, while the inner field of the third value is equal or greater than 1.
If the pattern does match, the inner field of the first value will be bound to the \lstinline|first| variable, and the execution branches into the \lstinline|if let| body.
This illustrates well, how the decision whether a pattern matches can entail a large number of conditions.

When used in \lstinline|match| expressions, the compiler also ensure that for every possible input value at least one pattern matches it; match-statements are guaranteed to be exhaustive.
Because of that, \autoref{lst:non_exhaustive} results in a compiler error, as the value \lstinline|1| is not handled by any of the patterns.

\begin{lstlisting}[label=lst:non_exhaustive,language=Rust,caption={Non-exhaustive \lstinline|match| expression}]
match number {
    0 => println!("zero"),
    2.. => println!("large number"),
}
\end{lstlisting}

The exhaustiveness check is especially useful when matching on enums, as it ensures all variants are handled (if there is no catch-all default condition).
Rust's \lstinline|match| statements have a simple control flow: always exactly one \lstinline|match| arm is executed~\footnote{if multiple patterns match, only the first patterns match-arm is executed}.
That is, compared to switch-statements in C/C++, where any number (none, one, multiple, all) of the \lstinline|case| blocks may be executed~\cite{ISO9899,ISO14882}. 

\subsection{Macro hygiene}
\label{sec:macro_hygiene}

Rust features two mechanisms for macros, both of which operate on the token stream (valid syntax items)~\citerustref{macros-by-example.html}.
A Rust macro in effect is a function, consuming a token stream and yielding a (potentially mutated) token stream, executed during compile time.
Compared to macros in other programming languages, like \lstinline|#define| in C and C++, Rust macros operate on the compiler \ac{AST} rather than the raw source code as text.

Operating on the \ac{AST} allows the compiler to implement \say{hygiene}~\citerustref{macros-by-example.html\#hygiene}: unless provided as macro parameters, identifiers defined by a macro live in a separate scope, and are not accessible by the code surrounding the macro expansion.
Hygiene also allows macros defined in a \textit{Rust edition}~\footnote{\url{https://doc.rust-lang.org/edition-guide/editions/}} (opt-in changes to Rust's syntax and semantics) to be expanded within code using a different \textit{edition}, with the code generated by the macro retaining the behavior of the macro's edition.

For example, in \autoref{lst:hygiene} the program will print 0, because the \lstinline|example| variable defined by the macro is different than the one defined by the caller, even though the identifier is the same.
One consequence of this is that macros cannot be expanded by an external tool, because the expanded source code would lack the hygiene information, possibly resulting in unintended program behavior change.

\begin{lstlisting}[label=lst:hygiene,caption=Hygienic Rust macro]
macro_rules! redefine {
    ($value:expr) => { let example = $value; }
}

let example = 0;
redefine!(10);
println!("{example}");
\end{lstlisting}

Macro hygiene thus prevents one of the ways instrumentation for code coverage is usually injected by industry tools.
It is not possible for a tool to pre-process the source code, expand the macros and then inject calls to the instrumentation runtime into the expanded code.
Macros are however vital to Rust, and prohibiting the use of macros is not feasible.
Emerging from that, a Rust \ac{MCDC} tool has to comprise large parts of a functional Rust compiler, or has to be implemented as a Rust compiler plugin. 

\section{Role of pattern-matching in Rust}
\label{sec:role_of_pattern_matching_in_rust}

Development of safety-critical software tends to resort to simple, well-understood techniques and language features, with coding standards like MISRA-C\footnote{\url{https://misra.org.uk/}} forbidding the use of features that are complex or could be misused. This begs the question, can we simply forbid the use of pattern-matching when writing Rust programs, avoiding all and any trouble with \ac{MCDC}?

Pattern-matching is core to the Rust language, and major language features rely on it. The most prominent of those features are enumerations (\lstinline|enum|, see \autoref{sec:enums_in_rust}), as pattern matching is the only supported way to access the inner fields of enum variants. Enums are widespread in the Rust ecosystem, including the core library: in Rust, nullability of fields and variables is expressed with the \lstinline|Option<T>|~\footnote{\url{https://doc.rust-lang.org/core/option/enum.Option.html}} enum (due to the intentional lack of \lstinline|null|), and error handling is done using the \lstinline|Result<T, E>|~\footnote{\url{https://doc.rust-lang.org/core/result/enum.Result.html}} enum.

Pattern-matching is also present in every variable assignment, as the left side of a \lstinline|let| statement is specified to be a pattern~\citerustref{statements.html\#let-statements}. For example, in \lstinline|let x = 42;|, \lstinline|x| is an \hyperref[sec:identifier_sub_pattern]{identifier sub-pattern}, not just an identifier.

Regardless of other language features requiring pattern-matching, we also believe that the benefits of pattern-matching in safety-critical software outweigh the complexity the feature introduces. Enums and \lstinline|match| allows representing state machines and data attached to each state, with the compiler enforcing all states to be handled (thanks to exhaustiveness) while state-specific data can only be accessed in its respective state. Having these checks performed by the compiler itself rather than a third-party static analyzer increases the confidence in the code during the development process.

Because of this, we argue that pattern-matching should not be forbidden when writing Rust, and thus its interactions with \ac{MCDC} should be explored and understood.

\section{\ac{MCDC} for Rust's pattern matching}
\label{sec:mcdc-for-pattern-matching}

As highlighted above, pattern matching does not fit the literal definition of decision, as boolean operations are not required for pattern matching.
Despite that, pattern matching influences both values and branching, often substitutes what would be boolean expressions in C, and is eventually compiled into conditions and branches.
It follows, that pattern-matching shall be considered as decisions.
To describe how and when a pattern becomes a decision, first we need to establish some language to talk about pattern matching in detail.

\subsection{Patterns as a tree of sub-patterns}
\label{sec:pattern_as_tree}

\begin{figure}
    \centering

\tikzset{
    onetoken/.style={},
	setunderscoreshift/.code={\pgfmathsetlengthmacro{\underscoreshift}{-.5*width("#1")-\pgfkeysvalueof{/pgf/inner xsep}}},
	underscoreshift/.style={xshift=\underscoreshift}
}

\begin{tikzpicture}
	\begin{scope}[every node/.style = token]
		\node (token_stream) {\lstinline|[_, _, _]|};

		\node[below left = of token_stream] (sub_1) {\lstinline|Some(_)|};
		\node (sub_2) at (sub_1-|token_stream) {\lstinline|Some(_)|};
		\node[below right = of token_stream] (sub_3) {\lstinline|rest @ _|};

		\node[below = of sub_1] (sub_sub_1) {\lstinline|var|};
		\node[below = of sub_2] (sub_sub_2) {\lstinline|2..=8|};
		\node[below = of sub_3] (sub_sub_3) {\lstinline|..|};
	\end{scope}

	\draw (token_stream.215) -- (sub_1.north);
	\draw (token_stream) -- (sub_2.north);
	\draw (token_stream.325) -- (sub_3.north);
	
	\draw (sub_sub_1) -- (sub_1);
	\draw (sub_sub_2) -- (sub_2);
	\draw (sub_sub_3) -- (sub_3);

 \end{tikzpicture}
    \caption{Decomposition of example pattern into sub-patterns. The \lstinline|_| denotes an abitrary inner value}
    \label{fig:decomposition_sub_patterns}
\end{figure}
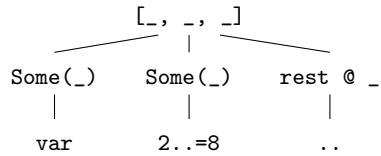

Fundamentally, pattern-matching is tied to the \ac{AST}, each pattern denotes a sub-tree of the \ac{AST}.
Patterns can nest, therefore one sub-tree of the \ac{AST} (representing a pattern) can contain further sub-sub-trees (each again representing a sub-pattern).
Anticipating the complexity patterns can reach, this view is useful, as it allows to consider properties of each sub-pattern individually.
The above is illustrated on the example of matching items of type \lstinline|&[Option<i32>]|:

\begin{lstlisting}[label=lst:complex-slice,caption=Complex slice pattern]
[Some(var), Some(2..=8), rest @ ..]
\end{lstlisting}

Splitting \autoref{lst:complex-slice} into sub-patterns, and arranging them into a tree with the outermost pattern as the root of the tree (which we will call the \textit{root sub-pattern}), we get the tree depicted in \autoref{fig:decomposition_sub_patterns}.
The rest of this paper will assume this view of patterns as a tree of sub-patterns, and we'll define \textit{child sub-patterns} as the sub-patterns directly below a given sub-pattern in the \ac{AST}

\subsection{Pattern refutability}
\label{sec:pattern_refutability}

As hinted in \autoref{sec:pattern_matching_intro}, patterns can, but do not necessarily implicate decisions.
One property useful to this distinction is \textit{refutability}~\citerustref{patterns.html\#refutability}.
A pattern is deemed to be refutable when it is possible for any value of the correct type not to be matched by the pattern, while it is deemed irrefutable when any possible value of the correct type will match the pattern.
Pattern refutability necessitates pattern matching to be treated as decision; matching of a refutable pattern implicates a conditional branch point.
On the contrary, matching an irrefutable pattern implies no decision and no branching.

For simple patterns it is possible to determine their refutability by just considering the kind of pattern: as an example, the wildcard pattern (\lstinline|_|) is irrefutable because it can by definition match any value, while a literal pattern (\lstinline|1|) is refutable because there are more numbers than just 1.
In other cases a pattern's refutability depends on the surrounding code: for example, the inclusive range pattern \lstinline|0..=255| is refutable if the matched type is a 32-bit integer, while it is irrefutable if the matched type is an 8-bit unsigned integer (as the range covers all possible values represented by that type).
There are also cases where a pattern's refutability depends on the refutability of its child sub-patterns: for example, a tuple pattern is refutable if any of the sub-patterns matching tuple fields are refutable, while it is irrefutable if all the sub-patterns themselves are irrefutable.

To aid the \ac{MCDC} analysis of patterns, for this paper we categorize sub-patterns into three cases:

\begin{itemize}
    \item Directly refutable sub-pattern: there are values of the correct type this sub-pattern does not match.
    \item Indirectly refutable sub-patterns: the sub-pattern itself matches all possible values of the correct type, but its direct child sub-patterns are refutable.
    \item Irrefutable sub-pattern: the sub-pattern matches all possible values of the correct type, and its direct child sub-patterns are irrefutable.
\end{itemize}

With this categorization of sub-patterns, we can define a pattern as refutable when the root sub-pattern is directly refutable or indirectly refutable, and we can define a pattern as irrefutable when the root sub-pattern is irrefutable.

\subsection{Refutability of sub-patterns}


\subsubsection{Literal sub-patterns}

Literal sub-patterns match any Rust literal, like booleans, \lstinline|char|s, bytes, strings, integers and floats.
They are always directly refutable, as none of the types of literal have only one possible value.

\subsubsection{Identifier sub-patterns}
\label{sec:identifier_sub_pattern}

Identifier sub-patterns bind the value they match to a local variable, optionally preceded by the \lstinline|ref| and \lstinline|mut| modifiers.
In this form they are always irrefutable.

Identifier sub-patterns can also be used to bind a matched sub-pattern to a local variable, with the \lstinline|VARIABLE @ SUB_PATTERN| form.
In this form they are irrefutable if the sub-pattern is irrefutable, otherwise they shall be considered indirectly refutable.

Note that it is only possible to determine whether a pattern is an identifier sub-pattern after performing semantic and type analysis of the surrounding code.
Otherwise determining whether a sub-pattern is an identifier pattern (binding to a variable) or a \hyperref[sec:path_sub_pattern]{path sub-pattern} is undecidable.

\subsubsection{Wildcard sub-patterns}

Wildcard sub-patterns match any possible value, and are written as \lstinline|_|.
They are by definition irrefutable.

\subsubsection{Rest sub-patterns}

Rest sub-patterns match zero or more elements in a variable-length pattern that haven not been matched before or after it, and are written as \lstinline|..| (two consecutive dots).
They are by definition irrefutable.

\subsubsection{Range sub-patterns}

Range sub-patterns match values in the specified range.
If the underlying type being matched is a fixed-width integer or a \lstinline|char|, the sub-pattern is irrefutable if the range covers all possible values. Otherwise the sub-pattern is directly refutable.

\subsubsection{Reference sub-patterns}

Reference sub-patterns dereference a type, allowing to match the type behind the reference, with the \lstinline|&SUB_PATTERN| or \lstinline|&mut SUB_PATTERN| syntax. 
They can be elided due to the match ergonomics feature\footnote{\url{https://rust-lang.github.io/rfcs/2005-match-ergonomics.html}}.
They are irrefutable if the sub-pattern is irrefutable, otherwise they shall be considered indirectly refutable.

\subsubsection{Struct sub-patterns}

Struct sub-patterns match one or more fields of a struct, or of an enum variant that uses the struct-like representation for fields. 
They allow binding fields to a local variable, and matching the content of fields using sub-patterns.

If the underlying type matched by the sub-pattern is a struct, it is irrefutable when all its sub-patterns are irrefutable, otherwise it is indirectly refutable.

If the underlying type matched by the sub-pattern is an enum, it is directly refutable if the enum has more than one variant, it is indirectly refutable if the enum has only one variant and any of its sub-patterns is refutable, otherwise it is irrefutable.

\subsubsection{Tuple struct sub-patterns}

Tuple struct sub-patterns match one or more fields of a struct using the tuple syntax\footnote{\lstinline|struct Foo(i32, String);|}, or of an enum variant that uses the tuple-like representation for fields.
They allow binding fields to a local variable, and matching the content of fields using sub-patterns.

It uses the same refutability rules as struct sub-patterns.

\subsubsection{Tuple sub-patterns}

Tuple sub-patterns match the fields of a tuple using sub-patterns.
They are irrefutable if all of its sub-patterns are irrefutable, otherwise they are indirectly refutable.

\subsubsection{Grouped sub-patterns}

Grouped sub-patterns are parenthesis used to define the precedence of the pattern syntax. For example, \lstinline|&0..=5| is ambiguous, as it could refer to both \lstinline|(&0)..=5| and \lstinline|&(0..=5)|.
Grouped sub-patterns allow to express the wanted sub-pattern using parenthesis. They are irrefutable if its sub-pattern is irrefutable, otherwise they are indirectly irrefutable.

\subsubsection{Slice sub-patterns}

Slice sub-patterns match the contents of a fixed-sized array or a dynamically sized slice, with the syntax \lstinline|[SUB_PATTERN_1, SUB_PATTERN_2, SUB_PATTERN_3]|, and can contain an arbitrary number of sub-patterns.

If the underlying type matched is a fixed-sized array, it is irrefutable when all its sub-patterns are irrefutable, otherwise it is indirectly refutable.

If the underlying type matched is a dynamically sized slice, it is directly refutable if none of its sub-patterns are range patterns, it is indirectly refutable if any of its sub-patterns are refutable, otherwise it is irrefutable. 

\subsubsection{Path sub-patterns}
\label{sec:path_sub_pattern}

Path sub-patterns can refer to constants, associated constants, structs without fields, or enum variants without fields.
If the item referred to is a constant or an associated constant, they are directly refutable.
If the item referred to is a struct, they are irrefutable.
If the item referred to is an enum variant, they are irrefutable if the enum has only one variant, otherwise they are directly refutable.

\subsubsection{Or sub-patterns}

Or sub-patterns allow to match one of two or more sub-patterns, with the form of \lstinline{SUB_PATTERN_1 | SUB_PATTERN_2}.
In general, an or pattern is indirectly refutable only if all of its sub-patterns are refutable, otherwise it is irrefutable.
This behavior is the opposite of the other kinds of sub-patterns, since any irrefutable sub-pattern makes the whole or pattern irrefutable.

Determining the refutability gets more complex if its refutable sub-patterns cover all possible values when or-ed together.
For example, let's take an enum with two variants \lstinline|A| and \lstinline|B|. According to the general definition, the pattern \lstinline{A | B} would be indirectly refutable, because both sub-patterns are refutable (path pattern for an enum with multiple variants).
In practice though the pattern is irrefutable, because the or sub-pattern covers all possible variants, and thus always matches.



\subsection{Conditions and decisions in a pattern}

Given the definition of refutability above, we can confirm that irrefutable patterns are not decisions, as the pattern always evaluates positively regardless of the value provided as input.

Patterns that are refutable are however decisions, as the pattern can evaluate positively or negatively depending on the value provided as input.
We can also note that only directly refutable sub-patterns do influence the outcome of the decision, as both indirectly refutable and irrefutable sub-patterns always match (ignoring their child sub-patterns) and shall be considered as a constant expression.
Because of this, we can consider all directly refutable sub-patterns as the conditions of the decision.

\section{Further \ac{MCDC} Considerations}
\label{sec:mcdc-further-considerations}

Pattern matching poses the big question with regard to \ac{MCDC} for Rust, while boolean expressions behave just like assumed by DO-178C, CAST-10, etc.
However some further open points remain.
This section therefore addresses other open points, in particular  nested conditionals and error handling with the question mark operator. 

\subsection{\ac{MCDC} for nested conditionals}\label{nested}

Compared to other languages like C and C++, in Rust most constructs that check expressions and branch based on them, like \lstinline|if| and \lstinline|match|, are itself expressions rather than just statements.
This means they can be used in every place that accepts expressions, like assigning to a variable:

\begin{lstlisting}[caption={\lstinline|if| expression assigned to a variable}]
let foo = if bar > 10 {
    "large"
} else {
    "small"
};
\end{lstlisting}

Crucially, conditions are also expressions, which for example allows nesting an \lstinline|if| inside of an \lstinline|if| condition:

\begin{lstlisting}[label=lst:nested-if,caption={\lstinline|if| expression used as a condition}]
if if foo > 10 { "large" } else { "small" } == "large" {
    println!("foo is big");
}
\end{lstlisting}

Code like \autoref{lst:nested-if} would never be written by a human and would never pass code review, because it is extremely hard to comprehend.
Still, it is common for code similar to that to be \textit{generated} by macros:

\begin{lstlisting}[label=lst:macro-in-if,caption=Macro used as a condition]
// Reimplementation of the standard matches! macro:
macro_rules! custom_matches {
    ($expr:expr, $pat:pat) => {
        match $expr {
            $pat => true,
            _ => false,
        }
    }
}
let value = Some(10);
if custom_matches!(value, Some(1..)) {
    println!("number is large than 1");        
}
\end{lstlisting}

\autoref{lst:macro-in-if} is well-formed Rust code that would pass code review, but still results in a \lstinline|match| inside of the \lstinline|if| condition after macros are expanded.
Because of that, we must consider how this affects \ac{MCDC} coverage.

According to CAST-10 \say{no matter how the logic is distributed across the system’s computer program and modules, \ac{MCDC} will address it}~\cite{cast10}.
We can thus extract the nested decision as its own separate decision, and treat its results as a single condition inside of the outer decision.

\subsection{\ac{MCDC} for the question mark operator}

Rust's question mark operator (\lstinline|?|) can be used on values of type \lstinline|Option<T>| and \lstinline|Result<T, E>|~\citerustref{expressions/operator-expr.html\#the-question-mark-operator}.
When the operator encounters an \lstinline|Option::Some| or a \lstinline|Result::Ok|, it evaluates to the inner value of those enum variants.
Otherwise it diverges and returns from the function respectively \lstinline|Option::None| and \lstinline|Result::Err|.
The compiler lowers this into a \lstinline|match| expression, like this:

\begin{lstlisting}[caption=Desugaring of the question mark operator]
match value {
    Some(inner) => inner,
    None => return None,
}
\end{lstlisting}

We can thus consider the question mark operator like a nested \lstinline|match|, and treat it according to \autoref{nested}.

\subsection{Constants in Rust}

The definition of \ac{MCDC} recommended per CAST-10~\cite{cast10} requires evaluation to true/false as well as independent effect on decision outcome to be demonstrated only for non-constant conditions.
Since constant conditions are out of scope for \ac{MCDC}, a significant effort can be spared.
That is, provided that a sound definition of \say{constant} can be found, which in the case of Rust is not trivial.

In Rust there are two kinds of items that should be considered constant for the purpose of \ac{MCDC} analysis:

\begin{itemize}
    \item Literal values like numbers, strings, chars and booleans.
    \item \lstinline|const|s and associated \lstinline|const|s.
\end{itemize}

There are more Rust constructs which intuitively appear constant though, like \lstinline|static|s and immutable \lstinline|let| bindings.
From a cursory glance, Rust's ownership system would prevent mutation of \lstinline|static|s and immutable \lstinline|let| bindings, as it is not possible to obtain a mutable reference to them.
Unfortunately it is still possible to mutate the contents of them at runtime, thanks to Rust's interior mutability, which allows mutating a subset of types using immutable references~\citerustref{interior-mutability.html}.
Because of this, it is not correct to treat them as constant.

We can instead consider \lstinline|const|s and associated \lstinline|const|s as constants because they do not reside in memory, and are instead inlined into the executable code wherever they are referenced.
It is thus not possible to mutate them at runtime, since there is no memory to mutate at all.

\section{Related Works}
\label{sec:related_works}

The literature on pattern matching in the context of \ac{MCDC} is sparse.
Development of airborne (and safety critical) software often gravitated towards imperative, procedural languages like Ada and C, while the concept of pattern matching emerged from the functional programming paradigm.
In consequence, safety critical software development was almost never exposed to pattern matching.
Despite that, we provide a list of at least adjacently relevant discussions found in the literature.

Wang, Jonquet and Chailloux published their tool \textit{zamcov}, which promises various coverage measurements on applications written in OCAML~\cite{nonintrusive_coverage_ocaml}.
In the work, structural coverage analysis is achieved without instrumentation of the application code.
Instead, the runtime environment (an OCAML bytecode interpeter) is instrumented to yield coverage information during execution.
While the article formalizes pattern-matching for structural coverage, \ac{MCDC} remains in the the future work section.
The provided \ac{MCDC} definition ties decisions to branch-points, \say{decision is the Boolean expression evaluated [...] to determine the branch to be executed."}~\cite{nonintrusive_coverage_ocaml}, a view that is to be reject according to CAST-10~\cite{cast10}.
Due to being publicized one year before the release of DO-178C~\cite{do178c} and DO-330~\cite{do330}, certification aspects are only discussed alongside their predecessor DO-178B~\cite{do178b}.
Consequent upon this, no discussion of \ac{TQL} or DO-330~\cite{do330} specifics is present in the work.

A continuation of the aforementioned work is found in Wang's thesis \say{Langages Applicatifs et Machines Abstraites pour la Couverture de Code Structurelle}~\cite{ocaml_filtrage_par_motifs}.
Again a strong connection between boolean expression coverage and branch points is tied: \say{Plusieurs critères de couverture de code structurelle existent. Dans cette thèse, nous en utilisons un certain nombre, notamment en ce qui concerne la couverture des expressions à valeurs booléennes. Ces expressions ont la particularité d’entraîner les décisions lors des branchements.}~\cite{ocaml_filtrage_par_motifs}, approximately \say{Several structural code coverage criteria exist. In this thesis, we use a number of them, particularly with regard to the coverage of boolean expressions. These have the particularity of driving branching decisions.}.
Most notable, this work dives into the semantics of pattern-matching (denoted \say{filtrage par motifs}~\cite{ocaml_filtrage_par_motifs}) in the context of \ac{MCDC}.
Similar to our assessment in \autoref{sec:mcdc-for-pattern-matching}, Wang concludes that pattern-matching does not appear as boolean expressions in the code, but still compiles down to decisions~\cite[pp.~147]{ocaml_filtrage_par_motifs}.
Two possible solutions are named: either, to just check the branching outcomes at the object code level~\cite[pp.~147]{ocaml_filtrage_par_motifs}.
Or, to trust the compiler that it handles pattern matching as it would handle equivalent \lstinline|if| statements~\cite[pp.~147]{ocaml_filtrage_par_motifs}.
This however does not answer the question for \ac{MCDC} in Rust.
On the one hand, while \lstinline|match| statements in Rust can be rewritten to various variations of \lstinline|if| statements such as \lstinline|if let|\footnote{\url{https://doc.rust-lang.org/book/ch06-03-if-let.html}}, these do still contain the same pattern matching.
On the other hand, even matching a single pattern may imply multiple conditions, hence \ac{MCDC} is not generally satisfied by just measuring if the pattern matches or not.

Taylor and Derrick describe the implementation of an \ac{MCDC} tool for Erlang/OTP~\footnote{\url{https://www.erlang.org/}}~\cite{smother}.
Considering that Erlang also features pattern-matching, their work is similar in goal to ours.
They argue that the majority of decisions in Erlang code are not expressed as boolean algebra but as pattern matching, an observation that matches our experience in Rust.
It seems the authors consider only branch points as decisions (a misconception described and rectified in CAST-10~\cite{cast10}), however no specific definition is provided in the paper.
The tool, \textit{smother}, relies on injection of instrumentation routines to the source code and run-time reflection to both find conditions and decisions.
Observed decision evaluations are then sent to dedicated observer processes.
As highlighted in \autoref{sec:macro_hygiene}, this approach of source-code based instrumentation is not feasible for Rust code.
Besides that, the authors model the pattern matching using evaluation trees, a representation similar to our proposal of (sub-) pattern trees outlined in \autoref{sec:pattern_as_tree} and illustrated in \autoref{fig:decomposition_sub_patterns}.


\section{Certification}
\label{sec:certification}

To understand the application of \ac{MCDC} in certification, first we inspect the requirement for \ac{MCDC}.
DO-178C requires test coverage with \ac{MCDC} to be achieved for level A software, and \ac{DC} for level B --- both with independence~\footnote{verification activity is not performed by the developer of the item being verified}~\cite[ch. 6.4.4.2, 6.4.4.3]{do178c}.
The process of satisfying coverage as per the \ac{MCDC} measure can be split into three steps.
As prerequisite, a test suite traceable to the requirements has to be implemented.

\begin{enumerate}
    \item The software item's source code is analyzed to identify all conditions and decisions.
    This step can be performed by a human or by a code analysis tool
    \item Test suite is executed with instrumentation in place to observe branching (statement coverage), decision outcomes (\ac{DC}) and/or conditions (\ac{MCDC}).
    To enable the observation of program-internal state during execution, tool\footnote{both hard- and software} support is likely required for practical reasons
    \item The observations from step 2 are evaluated alongside the analysis from step 1, to measure the achieved coverage.
    While this can be performed by a human, tool support may be preferable to avoid human error in the tedious work
\end{enumerate}

If the desired coverage is not reached, shortcomings in the test suite are addressed and the process repeats.

DO-178C joint with DO-330 provide a framework to assess the \ac{TQL}~\cite[ch. 12.2.2]{do178c}\cite[Appendix B]{do330}, ultimately determining the assurance objectives of a software-tool.
The key to that is the intended use of a tool; three tool usage criteria are distinguished.
For both steps mentioned above, criterion 2 applies: \say{A tool that automates verification process(es) and thus could fail to detect an error, and whose output is used to justify the elimination or reduction of [...] process(es)}~\cite[ch. 12.2.2b]{do178c}.
Prepared with this classification and the software's level, DO-178C then dictates the \ac{TQL}: If the software is of level A or level B, then \ac{TQL}-4, for all other (lower) levels only \ac{TQL}-5 is mandated~\cite[Table 12-1]{do178c}.
Since both \ac{MCDC} and \ac{DC} are only required for level A and B respectively, any \ac{MCDC} tool (be it for code analysis/step 1, coverage observation/step 2 or coverage measurement/step 3) has to satisfy the requirements of \ac{TQL}-4.

The current trend predicts that \ac{MCDC} functionality will be implemented within the Rust compiler \textit{rustc} itself.
This however does not constitute that the entire Rust compiler code base has to reach \ac{TQL}-4.
In particular, a tool that fulfills multiple functions, such as \textit{rustc} (being a mixture of compiler, static analysis- and maybe soon \ac{MCDC}-tool), it is acceptable to only qualify those functions that are used to eliminate, reduce or automate processes.
Prerequisite to that is, that protection from other functionality of the tool can be demonstrated~\cite[ch. 11.1]{do330}.
Rust provides an excellent foundation on isolating functionality robustly.
However, as explained for example in \autoref{sec:macro_hygiene}, even step 1 involves large parts of the compiler.
As current upstream efforts towards \ac{MCDC} support in \textit{rustc} aim to implement both step 1 and step 3 inside \textit{rustc} while incorporating some of LLVM's code coverage machinery for step 2, major parts of \textit{rustc} will have to reach TQL-4.

Hence we conclude that while a sizable amount of work remains, a workflow to achieve \ac{MCDC} of Rust programs and consequently Rust in level A software is feasible.

\section{Conclusion \& Outlook}
\label{sec:conclusion}

This paper provides in-depth considerations and rational for the behavior of a Rust \ac{MCDC} tool.
Measuring a Rust program's test suite coverage with \ac{MCDC} is theoretically possible.
However, doing so requires amendment of the existing term definitions \textit{condition} and \textit{decision}.

In particular, refutable patterns must be treated as decisions.
Decomposition into a tree of sub-patterns reveals the individual conditions comprising the top-level decision whether a pattern matches or no.
Only \lstinline|const| declared items and literals are qualified for exemption from the \ac{MCDC} criterion, variables declared via \lstinline|let| and/or \lstinline|static| must be treated as non-constant.

The implementation of a Rust \ac{MCDC} checker may be hindered by the Rust compilers macro hygiene, which prohibits simple macro expansion analysis and instrumentation purposes.
Further on, semantic analysis and type checking of the surrounding code is required to determine whether a pattern actually is refutable (and hence poses a decision) or not.
Therefore a Rust \ac{MCDC} tool has to be either integrated into a working Rust compiler, or has to at least correctly re-implement (or re-use) significant parts of a Rust compiler front-end.

The Rust Project is working on an implementation of \ac{MCDC} coverage instrumentation inside of the compiler itself, based on the theoretical groundwork laid out in this paper.

%
%
\bibliographystyle{splncs04}
\bibliography{bibliography}
\end{document}